\documentclass[11pt,a4paper,twoside]{report}
\usepackage{amsmath,amssymb}
\usepackage{color}
\usepackage[pdftex,plainpages=false,pdfpagelabels]{hyperref}
\usepackage{graphicx}
\usepackage{url}
\usepackage{calc}
\usepackage{natbib}
\usepackage[font=small,labelfont=bf]{caption}
\usepackage[outercaption]{sidecap}
\renewcommand{\baselinestretch}{1.2}
\renewcommand\baselinestretch{1.05}
\makeatletter

\def\@oddfoot{{\hfil\thepage\hfil }}
\renewcommand\section{\@startsection {section}{1}{\z@}%
                                     {-3.ex \@plus -1ex \@minus -.2ex}%
                                     {2.ex \@plus.2ex}%
                                     {\sffamily\Large\bfseries}}
\renewcommand\subsection{\@startsection{subsection}{2}{\z@}%
                                     {-3.ex\@plus -1ex \@minus -.2ex}%
                                     {1.ex \@plus .2ex}%
                                     {\sffamily\large\bfseries}}
\renewcommand\subsubsection{\@startsection{subsubsection}{3}{\z@}%
                                     {-1.ex\@plus -.2ex \@minus -.1ex}%
                                     {.1ex \@plus .1ex}%
                                     {\sffamily\normalsize\bfseries}}
\renewcommand\paragraph{\@startsection{paragraph}{4}{\z@}%
                                     {3.25ex \@plus1ex \@minus.2ex}%
                                     {-1em}%
                                     {\sffamily\normalsize\bfseries}}

\renewenvironment{thebibliography}[1]%
      {\begin{description}}{\end{description}}
\makeatother

\makeatletter
\renewcommand\@biblabel[1]{#1} 

\renewcommand\newblock{\hskip .07em\@plus.33em\@minus.07em}

\begin{document}

\begin{center}
\vspace*{-10mm}
{\bf\LARGE
Discussion on \textit{Competition for Spatial Statistics for Large Datasets}

\bigskip}
{Roman Flury$^{1}$, Reinhard Furrer$^{1, 2}$\\[2mm]

$^1$ Dept. of Mathematics, University of Zurich
$^2$ Dept. of Computational Science, University of Zurich}\\[2mm]

E-mail: roman.flury@math.uzh.ch, reinhard.furrer@math.uzh.ch
\end{center} 

\newpage

\renewcommand\baselinestretch{1.5}\rm
\begin{center}
\vspace*{-10mm}
{\bf\LARGE
Discussion on \textit{Competition for Spatial Statistics for Large Datasets}}
\end{center}

\section*{Abstract}
We discuss the experiences and results of the \textit{AppStatUZH} team's participation in the comprehensive and unbiased comparison of different spatial approximations conducted in the \textit{Competition for Spatial Statistics for Large Datasets}.
In each of the different sub-competitions, we estimated parameters of the covariance model based on a likelihood function and predicted missing observations with simple kriging.
We approximated the covariance model either with covariance tapering or a compactly supported Wendland covariance function.

\section{Introduction}\label{sec:1}
The goal of the \textit{Competition for Spatial Statistics for Large Datasets} was to compare established spatial statistics methods on data of scientifically relevant sizes.
Such comparisons are sparse in the literature and therefore, we participated in this competition with great interest and highly appreciated the invitation from Marc G.~Genton.
The remainder of this section provides a brief overview of the methodology used for our contribution and describes the computational infrastructure and respective software implementations.
In Section~\ref{sec:exp}, we focus on how we conducted the different sub-competitions, including a concise discussion of the results we submitted as members of the \textit{AppStatUZH} team.
We conclude with a short discussion in Section~\ref{sec:discussion}.

Our paradigm consisted of the assumption of an isotropic Gaussian random process.
We addressed the large datasets by different approximations of the covariance function and deliberate misspecification of the likelihood (termed approximate likelihood in the following).  

In sub-competition~1, we used covariance tapering to estimate the parameters of the tapered Mat\'ern covariance function.
The central idea of tapering is to exploit that numerous entries of the covariance matrix are near to yet distinct from zero.
These negligible entries are tapered to zero by multiplying the covariance function of interest by a compactly supported covariance function.
This multiplication yields the tapered covariance function, which is positive-definite and compactly supported.
Its compact support leads to the desired computational gains when sparse matrix algorithms are applied to efficiently compute the Cholesky factorization of the covariance matrix for estimation and prediction.
Based on the smoothness parameter $\nu$ of the observed field, a different taper function is chosen, for $0 < \nu \le 0.5$ a spherical, for $0.5 < \nu \le 1.5$ a $\text{Wendland}_1$ and for $1.5 < \nu \le 2.5$ a $\text{Wendland}_2$ covariance function.
This choice is based on in-fill asymptotics arguments to guarantee asymptotic efficiency of the simple kriging mean-squared prediction error.
Details of covariance tapering and respective asymptotic results can, for example, be found in~\cite{Furr:Gent:Nych:06,Kaufman:etal:08}.
It is important to note that these approximate likelihood estimates are biased.

In sub-competition~2, we chose a misspecification approach by using a Wendland covariance function directly instead of a tapered one. 
As shown in \cite{Bevi:Faou:Furr:Porc:19}, misspecification of a Mat\'ern with a generalized Wendland covariance with appropriate parameters choices leads to asymptotic efficiency. 
As an additional approximation, we used only integer-valued generalized Wendland functions ($\text{Wendland}_{1,2,3}$) to exploit their closed-form.
The choice thereof was based on the maximum value of the approximate likelihood function.
For the subsampling, we first retained all observations within the empirical $\gamma$ and $1-\gamma$ quantiles ($\gamma=0.0001,0.01$ for 2a and 2b respectively) and then drew a random subsample to maximize the approximate likelihood.
We repeated these steps several times and used the average as the final estimate.
Based on this model and respective estimates, we also used simple kriging for prediction.

The \texttt{R}~\citep{R} implementation of the covariance functions and the spatial statics approximations are all available in the sparse matrix packages \texttt{spam}~\citep{Furr:Sain:10,Furr:Flur:Gerb:20a} and \texttt{spam64}~\citep{Gerb:Mosi:Furr:16,Furr:Flur:Gerb:20b}.
The respective approximate likelihood functions were maximized with the \texttt{optimParallel} package~\citep{Gerb:Furr:19}.
We run our code serially or in parallel on a cluster of 4--9 nodes on a compute server with 8 x Intel Xeon 10C E7-2850 2.0 GHz and 2TB memory (80 cores).

\section{Results}\label{sec:exp}
In sub-competition~1, we discuss our results concerning the RMSEs since our primary focus was on making predictions using covariance tapering.
When applying covariance tapering, we had to make some initial decisions.
First, we chose an appropriate taper function for each field.
For this purpose, we estimated the smoothness parameter with an educated guess based on the empirical covariogram and the criteria described in Section~\ref{sec:1}.
When we compare the chosen taper functions with the true smoothness parameter (cf. Table~1 in \citealp{Huang:21}), we have only misspecified taper functions for fields containing a nugget effect. 
Considering the true smoothness parameters are near the margins of the decision criteria between the mentioned taper functions, the misspecified taper functions did not seem to have an effect on the RMSE.
Second, we chose an appropriate taper range.
The idea for this choice was to use the estimated effective range, which was based on the empirical covariogram.
Because the density of the tapered covariance matrix increases with increasing taper range, we worked with an upper limit of 0.3 for computational efficiency.
We started analyzing each data set with a random subsample of at least 10'000 locations and then increased the subsample size or adjusted the taper range until the estimates became stable or computational limits were exceeded.
As the estimates of the tapered Mat\'ern are biased estimates of the Mat\'ern covariance function, we presumed low performance concerning respective MMOMs and MLOEs; cf.~Table~S2 in \cite{Huang:21}.
It is natural but also crucial to use the same model for parameter estimation and prediction.
Hence, we have used the tapered Mat\'ern covariance function with estimates from~1a for prediction at the 10'000 locations.
Therefore, we included all observations available and solved one massive linear system for each dataset.
\begin{figure}
  \centerline{\includegraphics{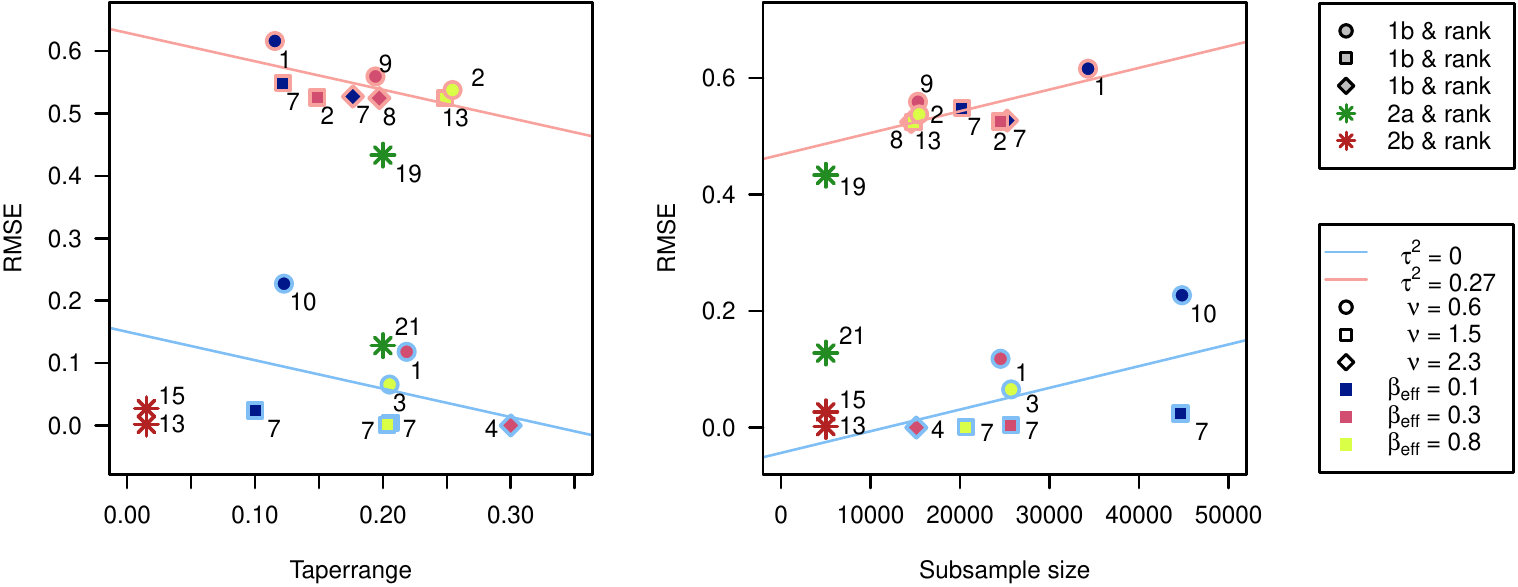}}
  \caption{\small RMSE's from prediction 1b, 2a \& 2b with respective ranks as function of the chosen taper range (left) and of the selected random subsample sizes (right).
  The straight lines show the average linear effect stratified for datasets from 1b containing a nugget effect to emphasize the respective correlations.}
  \label{fig:rmse}
\end{figure}
\cite{Huang:21} observed for other submissions that there is a distinct separation of RMSE's for fields with and without nugget effects.
This separation, evident in Figure~\ref{fig:rmse}, is in our results due to the simple kriging, which is essentially weighted spatial smoothing. 
Thus, the fields with a lower signal-to-noise ratio can be interpolated with smaller errors.
This inherent smoothing also explains the higher RMSEs for fields with a relatively small true smoothness $\nu = 0.6$; cf.~Figure~\ref{fig:rmse}, where our respective RMSEs are all above average linear trend.
Adherent to taper theory is that the RMSE of the predictions decreases with increasing taper range.
On the other hand, increasing the number of random subsamples seems to affect the RMSE in the right panel of Figure~\ref{fig:rmse} positively.
The positive effect may be relativized by the lower quality predictions for data set 3 (rank 10), which seems to have high leverage on this positive effect. 
Moreover, the fields with a small true effective range $\beta_{\text{eff}}$ seem to have a higher RMSE and distort the effect of the increasing subsamples.
Additionally, the distribution of the ranks in the $x\text{-direction}$ in both panels of Figure~\ref{fig:rmse} indicates no tendency regarding increasing taper range or subsample size.
It seems that, for these fields, the taper range should be chosen as large as the available memory allows for high-quality predictions.
This is supported by the fact that the number of neighbors is higher for a large taper range and a small subsample than for a large subsample.
Noteworthy in this context is that for datasets 4, 5, 12 and 13, the MLOE and MMOM are very large due to wide range estimates. 
Despite this, the predictions are still of high quality since we used the tapered covariance estimates that were jointly optimized.
Hence, the RMSE is reasonably low and we rank for these datasets in 1b as 4th, 7th, 8th and 2nd.

In sub-competition~2,  we first chose the type of compactly supported Wendland covariance function based on the approximate likelihood.
It turned out that smoothness equal to one, i.e.~$\text{Wendland}_1$ suited best for all four data sets.
Once this parameter was established, sill and nugget were estimated based on random subsamples of size 5'000.
Thereby, we set the ranges as 0.2 for 2a and 0.015 for 2b.
For computational efficiency, we used a narrower range in 2b, which has 900'000 observations.
Given that the fields in sub-competition 2a do not contain a nugget effect and are of equal size as the fields in 1b, we would expect only slightly higher RMSEs.
Though, the RMSEs of 2a are more than minimally higher than the average predictions from 1b, likely due to insufficient local taper ranges and not enough random subsamples.
In 2b, where we used narrower taper ranges, the RMSEs are much smaller and we rank higher among competitors.
Having noted that the total data sizes are considerably larger in 2b, the kriging predictions are directly more accurate.

\section{Discussion}\label{sec:discussion}
In our experience, maximum likelihood-based predictions are robust and competitive for large and huge datasets when combined with approximations.
Compared to the overall median prediction time in \cite{Huang:21}, covariance tapering is relatively slow.
However, it is a simple and intuitive approach with only a few tuning parameters that are highly robust and natural to control.
This competition demonstrates that covariance tapering delivers accurate spatial predictions, as we ranked fourth among strong competitors in sub-competition 1b.
Covariance tapering combined with \texttt{spam} and \texttt{spam64} already scales well for large data sets and naturally advances as computer technologies and \texttt{R} handling of long-vectors improve.

In case the process deviates from Gaussian regularity or stationarity, a careful choice of (computational) parameters can mitigate some of the model misspecification effects.
However, we deliberately used a Gaussian assumption as an approximation and were therefore aware that for some fields, we have a loss of efficiency compared to other approaches, e.g., including a Tukey g-and-h transformation.
However, the direct approximation of a spatial model using a compactly supported covariance function is relatively young and more experience needs to be gathered, e.g.~with simulation studies focusing on non-Gaussianity and non-stationarity.
Furthermore, improved numerical approximations would allow to efficiently model not only integer-valued generalized Wendland functions.

\section{Acknowledgment \& Contributions}
We thank Huang Huang and co-authors for organizing the \textit{Competition for Spatial Statistics for Large Datasets}, Matthew Heaton for inviting us to write this discussion and Dario Morciano and Carsten Rose for their IT support.
This work is supported by the Swiss National Science Foundation SNSF-175529.
\textit{AppStatUZH}: Roman Flury (1a,b, main), Federico Blasi (2a,b), Michael Hediger (1a,b), Stephan Hemri (2a,b) and Reinhard Furrer (2a,b, main; 1a,b).

\section{References}
{\small \bibliographystyle{mywiley}
\bibliography{discussion_bib.bib}}

\end{document}